\documentclass[runningheads]{llncs}
\usepackage{graphicx}
\usepackage{longtable}
\usepackage{multirow}
\usepackage{algorithm}
\usepackage[compatible]{algpseudocode}
\usepackage{enumitem}
\usepackage{hyperref}
\usepackage[caption=false]{subfig}

\newcommand{\ceil}[1]{\left\lceil #1 \right\rceil}

\begin{document}
\title{Approximate weighted 3-coloring\thanks{The study was carried out within the framework of the state contract of the Sobolev Institute of Mathematics (project FWNF-2022-0019)}}
\author{Adil Erzin\inst{1,2}\orcidID{0000-0002-2183-523X} \and Roman Plotnikov\inst{1}\orcidID{0000-0003-2038-5609} \and Georgii Zhukov\inst{2}}
\authorrunning{A. Erzin, R. Plotnikov, G. Zhukov}
\institute{Sobolev Institute of Mathematics, SB RAS, Novosibirsk 630090, Russia \and
Novosibirsk State University, Novosibirsk 630090, Russia\\
\email{adilerzin@math.nsc.ru}}

\maketitle              
\begin{abstract}
The paper considers the NP-hard graph vertex coloring problem, which differs from traditional problems in which it is required to color vertices with a given (or minimal) number of colors so that adjacent vertices have different colors. In the problem under consideration, a simple edge-weighted graph is given. It is required to color its vertices in 3 colors to minimize the total weight of monochromatic (one-color) edges, i.e. edges with the same colors of their end vertices. This problem is poorly investigated. Previously, we developed graph decomposition algorithms that, in particular, allowed us to construct lower bounds for the optimum, as well as several greedy algorithms. In this paper, several new approximation algorithms are proposed. Among them are variable neighborhood search, simulated annealing, genetic algorithm and graph clustering with further finding the optimal coloring in each cluster. A numerical experiment was conducted on random graphs, as well as on real communication graphs. The characteristics of the algorithms are presented both in tables and graphically. The developed algorithms have shown high efficiency.

\keywords{Edge-Weighted Graph \and Vertices 3-Coloring \and Approximation Algorithms \and Simulation}

\end{abstract}

\section{Introduction}
The paper \cite{Erzin:24}, referencing to the papers \cite{Acedo:15,Bandh:09,Chandra:21,Gui:18}, describes the importance of the problem considered in this paper. One application of the problem is the use of three different frequency bands in such a way as to minimize the penalty for using the same frequencies by cell towers covering a common area. If the importance of a link does not depend on the connection, then the edge weights can be set equal to 1. If some connection is more important than another, then the weight of the corresponding edge is higher. The problem is to color the vertices of the graph in 3 colors in such a way that the total weight of monochromatic edges, i.e. edges whose vertices are colored the same color, is minimal. This problem is relatively new for the discrete optimization community.

\subsection{Related Results}
Coloring of graph vertices is the assignment of a positive integer corresponding to the color number on each vertex of the graph. A \emph{proper} coloring is one in which adjacent vertices have different colors. The classical problem of vertex coloring is properly coloring the vertices of the graph using the minimum number of colors, which is called the chromatic number. A huge number of publications are devoted to this problem, an overview of which can be found, for example, in \cite{Borodin:13,Husfeldt:15,Kostochka:13,Thakare:24}.

In \cite{Al-Omari:06} and \cite{Mansuri:10} new heuristic graph coloring algorithms are proposed which are based on known heuristic algorithms. One is adaptation of the Largest Degree Ordering \cite{Avanthay:03} algorithm, and another one is an adaptation of the Saturation Degree Ordering \cite{Falkenauer:96} algorithm. The proposed algorithms outperform the original results. The authors of \cite{Musliu:13} present an automated algorithm selection method based on machine learning for the graph coloring problem. For this purpose, they identify 78 features for this problem and evaluate the performance of 6 state-of-the-art (meta)heuristics. They use the data to train several classification algorithms to predict a new instance, the algorithm with the highest expected performance. To improve the performance of the machine learning algorithms, they investigate the impact of parameters and evaluate different data discretization and feature selection methods. Finally, they evaluate their approach, which exploits the existing techniques and the automated algorithm selection, and compares it with existing heuristic algorithms. Experimental results show that the solver based on machine learning outperforms previous methods on benchmark instances. Several papers are devoted to the Borodin–Kostochka conjecture that in every graph with maximum degree $\Delta\leq 9$ chromatic number satisfies $\chi\leq\max\{\omega,\Delta-1\}$, where $\omega$ is the maximum clique size. Thus \cite{Cranston:15} proves that certain conjectures that are prima facie weaker than the Borodin–Kostochka conjecture are in fact equivalent to it. \cite{Choi:23} proves the analogous result of the Borodin–Kostochka conjecture for list coloring: $\chi<\Delta$ for all graphs with $\Delta$ sufficiently large and $\omega<\Delta$. \cite{Maus:23} presents a deterministic algorithm to compute an $O(k\Delta)$-vertex coloring in $O(\Delta/k)+\log n$ rounds, where $k\in [1,O(\Delta)]$ can be freely chosen. The authors of \cite{Mosawi:24} propose a new approximation algorithm for finding an approximation of a chromatic number by using a graph adjacency matrix to colorize or separate a graph. They provide some examples to compare the performance of their algorithm to other available methods. They make use of the Dolan-Mor\.{e} performance profiles to assess the performance of the numerical algorithms to compare the efficiency of the proposed approach in comparison with some existing methods. \cite{Eppstein:03} proposes an $O((4/3+3^{4/3}/4)^{n})$-time algorithm for calculating the chromatic number.

The problem of coloring a planar graph with 3 colors is polynomial solvable if the graph contains 3 or fewer triangles, and NP-hard in the general case \cite{Cavallaro:21,Dailey:80}. If the planar graph is triangles-free, there is an $O(n)$-time algorithm to color it with 3 colors \cite{Dvorak:9}, where $n$ is the number of nodes. The coloring of 3-colorable graph can be done with $O(1.3217^{n})$ time complexity \cite{Meijer:23}. The planar graphs are 4-colorable \cite{Gonthier:8}, but there is no any polynomial time algorithm that constructs such coloring. However, there is a linear 5-coloring algorithm for a planar graph \cite{Chiba:81}. In \cite{DeIta:12} some necessary conditions are described for 3-coloring.

Several papers are devoted to the problem under consideration. \cite{Gui:18} presents the considered in this paper problem as a boolean quadratic programming and proposes a greedy algorithm for its solution. The authors conducted a simulation in which their greedy algorithm was compared to the algorithm described in \cite{Bandh:09}. \cite{Chandra:21} uses a genetic algorithm to solve the problem. In \cite{Abdullah:14}, a graph coloring algorithm based on an incidence matrix is proposed. \cite{Chrost:13} presents an algorithm that allows multi-criteria optimization in different network deployment scenarios to minimize the number of conflicts.

\section{Our Contribution}
In our previous paper \cite{Erzin:24}, we state the problem as a boolean linear programming, as well as a boolean quadratic programming. We use loss-less partitioning of the graph into connected subgraphs. Then, for still large resulting subgraphs, we propose several decomposition algorithms to construct connected components of ``small'' sizes. After such clustering, we temporarily remove the edges between some components. Using the mathematical formulations of the problem, for each component, we build an optimal solution by applying the CPLEX package. This made us possible to find a non-trivial lower bound for the optimum. We also proposed different greedy algorithms and local search procedures, which were tested both on random and industrial instances.

In this paper, we use a spectral clustering approach for the graph decomposition and propose several new approximation algorithms based on well-known meta-heuristics: simulated annealing, variable neighborhood search, and genetic algorithm. We also propose a novel method that iteratively improves solution by solving exactly the problem on subgraph. We implemented all the proposed algorithms and performed extensive numerical experiment both on the randomly generated and industrial cases. The conducted experiment shows a high efficiency of the proposed methods and allows us to define the best approach for each test instance.

The rest of the paper is organized as follows. Section 3 defines the problem. In Section 4, a novel approach to graph decomposition is presented. New approximation algorithms are described in Section 5. Section 6 contains the description results and analysis of an experimental study. Section 7 concludes the paper.

\section{Problem Formulation}
Let us have a simple undirected edge-weighted graph $G=(V,E)$ where $V$ is the set of vertices ($|V|=n$) and $E$ is the set of edges. To each edge $(i,j)\in E$ a weight $a_{ij}$ is assigned. We call the edge $(i,j)$ a \emph{single-color} if the incident vertices $i$ and $j$ have the same color.

\textbf{The problem is to color the vertices of the graph $G$ using 3 colors such that the sum of weights of single-color edges is minimal.}

Since the problem of 3-coloring of a planar 4-regular graph is NP-complete \cite{Dailey:80}, then the problem under consideration is NP-hard.

We proposed the boolean quadratic programming (BQP) formulation for the problem in \cite{Erzin:24}, which is a special case of the model (I) from \cite{Gui:18}. For the convenience of the reader, we restate the problem. To do this, we introduce the following variables:

$$
 x_{i}^{k} = \left\{
  \begin{array}{ll}
    1,\ if\ vertex\ i\ has\ color\ k;\\
    0,\ otherwise.
  \end{array}
\right.
$$
where $i=1,\ldots,n;\ k=1,2,3$.

Then the problem is as follows.
\begin{equation}\label{e1}
  \sum_{j=1}^n\sum_{k=1}^{3} a_{ij}x_{i}^{k}x_{j}^{k}\to\min_{x_{i}^{k}\in\{0,1\}};
\end{equation}
\begin{equation}\label{e2}
  \sum_{k=1}^{3} x_{i}^{k}=1, \ i=1,\ldots,n.
\end{equation}

In \cite{Erzin:24}, we proposed also a boolean linear programming (BLP) formulation of the problem. Let us introduced the additional variables:

$$
 y_{ij}^{k} = \left\{
  \begin{array}{ll}
    1,\ if\ both\ vertices\ of\ the\ edge\ (i,j)\in E\ have\ color\ k;\\
    0,\ otherwise.
  \end{array}
\right.
$$

Then the BLP is as follows.
\begin{equation}\label{e3}
  \sum_{(i,j)\in E} \sum_{k=1}^{3} a_{ij}y_{ij}^{k} \to\min_{x_{i}^{k}, \ y_{ij}^{k} \in \{0, 1\}};
\end{equation}
\begin{equation}\label{e4}
  \sum_{k = 1}^{3} x_{i}^{k}=1,\; i=1,\ldots,n;
\end{equation}
\begin{equation}\label{e5}
  y_{ij}^{k}\geq x_{i}^{k}+x_{j}^{k}-1,\ (i,j)\in E, \; k=1,2,3.
\end{equation}

After continuous relaxation of the (\ref{e3})-(\ref{e5}), a zero value for the functional (\ref{e3}) is achieved. Therefore, it is not possible to derive a non-trivial lower bound for the optimum.

Both small-dimensional problems (\ref{e1})-(\ref{e2}) and (\ref{e3})-(\ref{e5}) can be solved using CPLEX or Gurobi solvers.

\section{Graph Decomposition}
Using decomposition, we split the graph into relatively small components, and then solve the problem in each component separately and combine the solutions to get a coloring for the entire graph.

In \cite{Erzin:24}, we described the loss-less decomposition of the graph to split it into connected components, and presented several methods for graph clustering, when some edges were temporarily deleted. First clustering algorithm is based on consistently selecting ``heavy’’ edges. The second one is based on selecting random edges. The third algorithm is based on a random selection of vertices and heavy edges. Using these procedures described in \cite{Erzin:24}, a lower bound can be calculated using the optimal coloring of each resulting connected component (cluster).

In this paper, we adapt the spectral clustering algorithm \cite{Luxburg:06} and compare it with those proposed in \cite{Erzin:24}.

One of the main result of decomposition is to get a tight lower bound to the objective function of the original problem. To do this, we want to make the total weight of the edges within each cluster high and between the nodes of different clusters — low. Spectral clustering has recently become truly popular. It yields good quality results and outperforms many other algorithms. In this paper, we consider the Normalized spectral clustering, according to Ng, Jordan, and Weiss \cite{Jordan:01}.

Let us describe this algorithm. We are given an undirected simple graph $G = (V, E)$ with $n=|V|$ nodes. A non-negative weight $a_{ij}$ is assigned to each edge $(i, j) \in E$. We need to split the nodes $V$ it into $k$ clusters with the properties described above. Denote by $d_{i} = \sum_{j=1}^{n} a_{ij}$. Let $D$ be a square matrix of size $n$ with elements $d_{i}$ on the diagonal and zeros in the remaining positions. The normalized graph Laplacian is $L_{sym} = I - D^{-1/2}WD^{-1/2}$.

Algorithm \cite{Luxburg:06}:
\begin{itemize}[itemsep=1em]
    \item Compute the normalized Laplacian $L_{sym}$.
    \item Compute the first $k$ eigenvectors $u_{1}$,..., $u_{k}$ of $L_{sym}$, corresponding to the $k$ smallest eigenvalues.
    \item Let $U \in R^{n \times k}$ be the matrix containing the vectors $u_{1}$,..., $u_{k}$ as columns.
    \item Form the matrix $T \in R^{n \times k}$
    from $U$ by normalizing the rows to norm 1,
    that is set $t_{ij} = u_{ij} / (\sum_{k}^{} u_{ik}^{2}) ^ {1/2}$
    \item For $i =$ 1,..., $n$, let $y_{i} \in R^{k}$ be the vector corresponding to the $i$-th row of $T$.
    \item Cluster the points $(y_{i})_{i=1,...,n}$ with the $k$-means algorithm into $k$ clusters.
\end{itemize}

The final step of the algorithm involves $k$-means algorithm \cite{MacQueen:67}. However, if the $k$-means algorithm is applied promptly, the resulting clusters may not be balanced in terms of size. There will be both big and small clusters. In order to run CPLEX on the resulting clusters, it is necessary to limit the number of nodes in each cluster (CPLEX solves the problem in a reasonable time if the number of vertices is less than 100). There are variations in the $k$-means algorithm implementation that consider restrictions on the minimum and maximum cluster size. One such variation can be found in \cite{Bradley:00}. However, we will use a different approach and modify the $k$-means algorithm to meet our specific requirements.

Assume we want to have the number of vertices in each cluster not more than $Q$. Then we set $k = \ceil{n / Q}$. Initialize the cluster centers with randomly generated vectors drawn from a multidimensional uniform distribution on $[0, 1]^{k}$. Until the maximum number of iterations is reached or the clusters do not change, perform a step of the classical $k$-means algorithm. For clusters containing more than $Q$ vertices, select the vertex $v$ with the lowest weight within the current cluster (the weight of a vertex in a cluster is the total weights of the edges incident to it in this cluster). Remove $v$ from the cluster. Among all clusters with the number of vertices less than $Q$, we find one in which $v$ has the highest weight and adds it to this cluster. Repeat the described procedure until the cluster size is greater than $Q$. Use the traditional $k$-means algorithm to update the center points.

The pseudocode of this algorithm can be found in Algorithm \ref{SC}. Let us describe the used functions:
\begin{itemize}[itemsep=1em]
    \item \textit{GetT(W, k)} returns matrix $T$ described in the algorithm above.
    \item If the clusters are NULL, \textit{GetCenters(T, clusters, n, k)} returns a matrix of size $n \times k$ with random numbers drawn from a uniform distribution over the interval [0, 1]. Otherwise, it returns the centers according to the $k$-means algorithm.
    \item \textit{GetClusters(newCenters, T)} returns the clusters according to the $k$-means algorithm.
    \item \textit {GetMinVertex(cluster, W)} returns the vertex $v$ with the minimum total weight of the edges incident at $v$ in \textit{cluster}.
    \item \textit {GetCluster(clusters, W, v, Q)} returns the cluster number in which the vertex $v$ has the highest total weight of edges incident at it among all clusters with fewer than $Q$ vertices.
    \item \textit{IsChanged(clusters, newClusters)} returns \textit{FALSE} if \textit{clusters} equals \textit{newClusters}, otherwise returns \textit{TRUE}.
\end{itemize}

We will show in the Simulation section that such decomposition technique frequently yields more tight lower bound than the techniques described in \cite{Erzin:24}. However, not in all cases, therefore, the methods described in \cite{Erzin:24} can be used as well.

\begin{algorithm}[!hbtp]
\begin{algorithmic}[1]
\STATE \emph{Input}: $W$ (weight matrix of the graph $G$), $n$ (number of vertices), $k$ (number of clusters), $maxIter$ (maximum number of iterations), $Q$ (maximum size of cluster);
\STATE \emph{Output}: clusters (list of clusters);
\STATE $centers \: \leftarrow \: NULL$;
\STATE $clusters \: \leftarrow \: NULL$;
\STATE $T \: \leftarrow \: GetT(W, \: k)$;
\WHILE{$maxIter \: > \: 0$}
  \STATE $maxIter \: -= \: 1$;
  \STATE $newCenters \: \leftarrow \: GetCenters(T, \: clusters, \: n, \: k)$;
  \STATE $newClusters \: \leftarrow \:  GetClusters(newCenters, \: T)$;
  \STATE $count \: \leftarrow \: 0$;
  \WHILE{$count \: < \: newClusters.size()$}
    \IF {$newClusters[count].size() \: \leq \: Q$}
      \STATE $continue$;
    \ENDIF
    \WHILE {$newClusters[count].size() \: > \: Q$}
      \STATE $v \: \leftarrow \: GetMinVertex(newClusters[count], \: W)$;
      \STATE $numCluster \: \leftarrow \:  GetCluster(newClusters, \: W, \: v, \: Q)$;
      \STATE $newClusters[count].erase(v)$;
      \STATE $newClusters[numCluster].insert(v)$;
    \ENDWHILE
    \STATE $count \: += \: 1$;
  \ENDWHILE
  \IF {$IsChanged(clusters, \: newClusters) \: = \: FALSE$}
    \STATE $break$;
  \ENDIF
  \STATE $centers \: \leftarrow \: newCenters$;
  \STATE $clusters \: \leftarrow \: newClusters$;
\ENDWHILE
\STATE \textbf{return:} $clusters$;
\end{algorithmic}
\caption{Spectral Clustering} \label{SC}
\end{algorithm}

\section{Approximation Algorithms}

In this section, we describe new approximation algorithms, most of which are based on well-known meta-heuristic approaches. In \cite{Erzin:24} we proposed several constructive heuristics, local search procedures, and VND based method. Some of them are used in the algorithms proposed below.

\subsection{Iterative Partial Improvements}
First, we describe the iterative partial improvements (IPI) that start with some initial solution and perform the iterative improvements by solving the problem on the small-sized subgraphs. In each iteration, we apply CPLEX to find an optimal coloring for the subgraph $G^{\prime} = (V^{\prime}, E^{\prime})$ with a fixed set of vertex colors in $V \setminus V^{\prime}$. Assume we want to limit the number of vertices in a subgraph to $n^{\prime}$. Let us describe the algorithm. \\

\begin{enumerate}[itemsep=1em]

    \item Create a set of subgraphs $S$. Initially, it is empty. Let $U \: = \: V$ be the set of vertices that have not yet been visited;
    \item If $U$ is empty, go to point 6.
    Choose a random vertex $v$ from $U$.
    Let $V^{\prime}$ be a set of one vertex $v$. Remove $v$ from $U$. Let $N$ be the set of vertices from $U$ adjacent to some vertices in $V^{\prime}$ in $G$;
    \item If the size of $V^{\prime}$ equals $n^{\prime}$ or $N$ is empty, go to point 5;
    \item Select a vertex $u$ in $N$ with a probability proportional to its weight in a subgraph induced by a set of vertices $V^{\prime}$ (the vertex weight is the sum of the weights of the incident edges). Add $u$ to $V^{\prime}$. Remove $u$ from $U$ and from $N$. Add to $N$ all vertices in $U \setminus N$ adjacent to $u$ in $G$. Go to point 3;
    \item Let $G^{\prime}$ be a subgraph induced by a set of vertices $V^{\prime}$. Add $G^{\prime}$ to $S$. Remove all edges of $G^{\prime}$ from $G$. Remove all the boundary vertices of the graph $G^{\prime}$ from $U$ (boundary vertex of the graph $G^{\prime}$ is a vertex in $G \setminus G^{\prime}$ adjacent to at least one vertex of $G^{\prime}$). Go to point 2;
    \item For each subgraph $\in$ $S$, we find an optimal coloring, assuming that the colors of its boundary points are fixed;
	\item If stop condition is not met, then go to point 1. Otherwise, stop calculating.
\end{enumerate}

We set the maximum component size to 60. Greater values lead to too long calculations by CPLEX.

\subsection{Hybrid Simulated Annealing}
Another algorithm that we propose is hybrid simulated annealing (HSA). Our algorithm is a modification of the classic simulated annealing algorithm \cite{Kirkpatrick:83}. The first modification is that we apply VND from \cite{Erzin:24} in each iteration. The second modification is that a $T_{0} \: / \: log(i +1)$ function is used instead of a linear function to change the temperature. This temperature change function decreases more slowly than the linear one and makes it possible to slow down the process of random searching. As a result, we can find a better solution than if we use a linear function. We restart the algorithm while the stop condition is not satisfied. We select the following parameters for the algorithm: the initial temperature = 100, the minimum temperature = 0.\\

\subsection{Variable Neighborhood Search}
Our implementation of the variable neighborhood search algorithm is similar to the extension of basic VNS from \cite{Hansen:08} (algorithm 12). Its pseudocode can be found in the Algorithm \ref{VNS}. However, there are a few distinctions. The number of iterations in the nested loop depends not only on $k_{max}$, that corresponds to the maximum number of movements in the neighborhood system of shaking operator, but also on the new parameter $l_{max}$ that corresponds to the size of a neighborhood of shaking. The \textit{Shake} procedure is designed as follows.

Repeat $k$ times:

\begin{enumerate} [itemsep=1em]
\item Select a random vertex and build a breadth first search tree of height \textit{l} with the root at that vertex;
\item If the number of vertices in the resulting tree equals $n$, then the execution stops.
\item Generate a random number $s$ from \{1, 2\};
\item Perform shifting of the vertices in the resulting tree by the value $s$ mod 3;
\end{enumerate}
We use VND from \cite{Erzin:24} as the \textit{FirstImprovement} function. We experimentally defined the following parameters for the algorithm: $k_{max}$ = 10, $l_{max}$ = 50.\\

\begin{algorithm}[!hbtp]
\begin{algorithmic}[1]
\STATE \emph{Input}: $x$ (the initial solution), $k_{max}$;
\WHILE{Stop condition is not met}
    \STATE $k \: \leftarrow \: 1$;
    \REPEAT
        \FOR{$l \: = \: 1 $ \textbf{to} $k$}
            \STATE $x^{\prime} \: \leftarrow \: Shake(x, k, l)$;
            \STATE $x^{\prime \prime} \: \leftarrow \: FirstImprovement(x^{\prime})$;
            \STATE $x \: \leftarrow \: Best(x, \: x^{\prime \prime})$;
        \ENDFOR
    \UNTIL {$k \: = \: k_{max}$}
\ENDWHILE
\end{algorithmic}
\caption{VNS from \cite{Hansen:08}} \label{VNS}
\end{algorithm}

\subsection{Genetic Local Search}
The general scheme of genetic local search (GLS) is presented in the Algorithm \ref{GLS}. There are 4 parameters in this algorithm: $n_P$ is the number of solutions in population,  $n_R$ is the size of offspring in each iteration, $P_M$ is the probability of random mutation, $P_{VND}$ is the probability of improvement of solution by $VND$ from \cite{Erzin:24}. We also use the following notations: $P$ is the population of solutions (initial graph colorings); $F$ is the array of fitness values (the better solution, the greater fitness); $Q$ are the parents, i.e., a set of pairs of solutions that will produce an offspring; $R$ is the set of newly generated solutions (offspring).

Below, we describe each of the main steps of this algorithm.

\begin{itemize}
	\item \textbf{Initial population}. To have the initial population, we run all constructive heuristics from \cite{Erzin:24} and then apply the VND algorithm to the resulting solutions. In order to supplement the population with the required number of elements, we also generate solutions by the randomized version of the G4 heuristic from \cite{Erzin:24}. At each step, this algorithm considers a set of vertices with equal sums of adjacent edge weights and randomly selects one.
	\item \textbf{Fitness calculation}. As fitness value, we use inverted objective value multiplied by the constant normalization coefficient that allows us to avoid too small values when the edges weights are large.
	\item \textbf{Parents selection}. The parents are selected randomly with probabilities proportional to the fitness. Each pair is two different solutions, and each solution can be a part in different pairs (polygamy is allowed).
	\item \textbf{Crossover}. For the crossover, each parent solution is represented as an array of colors that corresponds to the vertices’ indices. Then, the one-point strategy is used to get two new solutions for a pair of parents. The index $i^* \in \{2, \dots, n\}$ is chosen randomly, and two new solutions are generated from the parents: in first solution, the colors of vertices 0, \dots, $i-1$ are the same as in the first parent, and the vertices $i$, \dots, $n$ are the same as in second parent, and in the second solution, the colors of vertices 0, \dots, $i-1$ are the same as in the second parent, and the vertices $i$, \dots, $n$ are the same as in the first parent.
	\item \textbf{Mutation}. Any offspring can mutate with a probability of $P_M$. We choose randomly a center vertex $v$ and a quantity of movements $k$ from $\left[\min(10, n/10), n/5 \right]$. After that, the colors of $k$ first vertices within breadth-first search starting at $v$ are randomly changed with a probability of 0.5.
	\item \textbf{Improvements}.  We apply to each offspring the VND algorithm from \cite{Erzin:24} with probability $P_{VND}$.
	\item \textbf{Join}. From $P \cup R$, the $n_P$ solutions with the highest fitness values are selected for the next population.
\end{itemize}

\begin{algorithm}[!hbtp]
\begin{algorithmic}[1]
\STATE \emph{Input}: $G$, $n_P$, $n_R$, $P_M$, $P_{VND}$;
\STATE $P \: \leftarrow \: GenerateInitialPopulation(n_P)$;
\WHILE {Stop condition is not met}
    \STATE $F \: \leftarrow \: CalculateFitness(P)$;
		\STATE $Q \: \leftarrow \: SelectParents(P, F, n_R)$;
		\STATE $R \: \leftarrow \: Crossover(Q)$;
		\STATE $R \: \leftarrow \: PerformMutations(R)$;
		\STATE $R \: \leftarrow \: PerformImprovements(R)$;
    \STATE $F_R \: \leftarrow \: CalculateFitness(R)$;
		\STATE $P \: \leftarrow \: Join(R, P, F, F_R)$;
\ENDWHILE
\STATE \textbf{return} the best solution of $P$;
\end{algorithmic}
\caption{Genetic local search (GLS)} \label{GLS}
\end{algorithm}

The procedures of crossover, mutation, and improvements are run in parallel. Each time when a new solution is generated, it is checked that it is not equal to any of existing solutions. We experimentally defined the following parameters for the algorithm: $n_P = 50$, $n_R = 25$, $P_M = 0.2$, and $P_{VND} = 0.8$.

\subsection{Combined Meta-heuristics Algorithm}
Finally, the method \textit{AllMetaheuristics} (AllMH) executes all previously described algorithms consecutively in the loop. The pseudocode of this method can be found in the Algorithm \ref{AllMetaheuristics}. In this method, the above meta-heuristics are used as subroutines, and in order to avoid getting stuck in one of them we used 5 times smaller time limit and maximum number of iterations without update of incumbent solution. At each iteration, single solution based methods HSA, VNS, and IPI start with current the incumbent solution, and GLS adds the current incumbent to the initial population instead of the worst from the generated ones.

\begin{algorithm}[!hbtp]
\begin{algorithmic}[1]
\STATE \emph{Input}: $G$ --- initial graph, $C$ --- initial solution;
\WHILE {stop condition is not met}
		\STATE $C \: \leftarrow \: HSA(C, G)$;
		\STATE $C \: \leftarrow \: VNS(C, G)$;
		\STATE $C \: \leftarrow \: IPI(C, G)$;
		\STATE $C \: \leftarrow \: GLS(C, G)$;
\ENDWHILE
\STATE \textbf{return} $C$;
\end{algorithmic}
\caption{AllMetaheuristics (AllMH)} \label{AllMetaheuristics}
\end{algorithm}

\section{Simulation}

All algorithms were coded in C++ for experimental evaluation and launched on i3-9100 CPU 8Gb RAM. The numerical testing involves three distinct categories of test instances. The first group includes real-world wireless network topologies. We were provided by several examples from a company that designs and configures large wireless communication networks. The coloring problem arises when it is necessary to minimize the PCI mod 3 interference. In such networks, the weight of each edge reflects the significance of the conflict. For the second group, edges between vertex pairs are randomly added with uniform distribution until reaching predefined edge counts, and the weight of each edge is a random uniformly distributed value on the interval $\left[0 ,100 \right]$. The third group handles unit disk graphs created by distributing points uniformly across a unit square and connecting vertices within specified radii, with edge weights corresponding to Euclidean distances. To unify conditions, we tried to maintain the same density $m \approx 10n$ for all generated graphs. For unit disk graphs, for this purpose, we chose an appropriate value of radius $r$ for each number of vertices. All the test instances and found solutions are publicly available by the following link: \url{https://disk.yandex.ru/d/mj5K_KK9aVXfzQ}.

We launched each of the randomized algorithms 10 times in the same instance. As a stop criteria, we used the following two conditions: (i) time limit of 10 minutes, (ii) 100 runs in a row without improvements. Each algorithm stopped as soon as at least one of these conditions is met. For small-sized problems, CPLEX was used to solve ILP model of the problem. For the larger sizes, we calculated two lower bounds: LB — the lower bound constructed by the algorithm described in \cite{Erzin:24}, LB2 — the lower bound constructed by the spectral clustering approach that is proposed in this paper. In order to estimate the improvement of each tested algorithm, we compared our results over the algorithms from \cite{Erzin:24}: the best greedy heuristic (Gr) and variable neighborhood descent applied to the result of Gr (VND).

\begin{figure}[!hbtp]
\centering
\subfloat[\label{tableInd1}Average objective and lower bound values]{\includegraphics[width=1\textwidth]{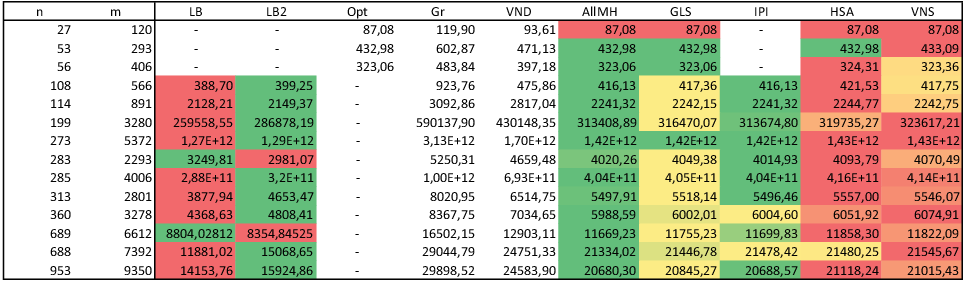}} \hfill
\subfloat[\label{tableInd2}Ratio estimations]{\includegraphics[width=1\textwidth]{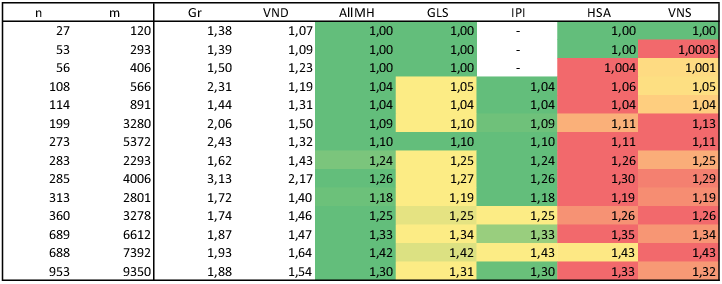}} \hfill
\caption{Industrial cases results} \label{tableInd}
\end{figure}

The results for industrial cases are presented in Fig. \ref{tableInd}. In the table at Fig. \ref{tableInd1}, the average value of each proposed algorithm is given for each tested instance. In this table, as well as in the following ones, cells are colored in range from green to red: the \emph{better} value (less -- for the objective function, greater -- for the lower bound) is colored in green, the worse one is colored in red, the middle values are colored in yellow or orange. The table with ratios is given in Fig. \ref{tableInd2}, the values presented there are strictly followed from Fig. \ref{tableInd1}, they are calculated by division of objective values and the maximum of lower bounds or optimal value when it was calculated. In a significant number of cases, IPI and GLS outperform other algorithms, but the difference with the worst result is not so significant, it never exceeds 4$\%$. The ratio of the best result is always less than 1.43, and for the smaller size, it is not greater than 1.1. These results show a high quality not only of the algorithms but also of lower bound. The lower bound LB2 yielded by spectral clustering approach is better than LB in almost all cases. For small-size problems ($n < 100$), when optimal solution was found by CPLEX, all meta-heuristics yield either optimal solution or solution that differs from optimum within 0.5$\%$.

\begin{figure}[!hbtp]
\centering
\subfloat[\label{graphicInd1}$n = 688,\:m=7392$]{\includegraphics[width=0.47\textwidth]{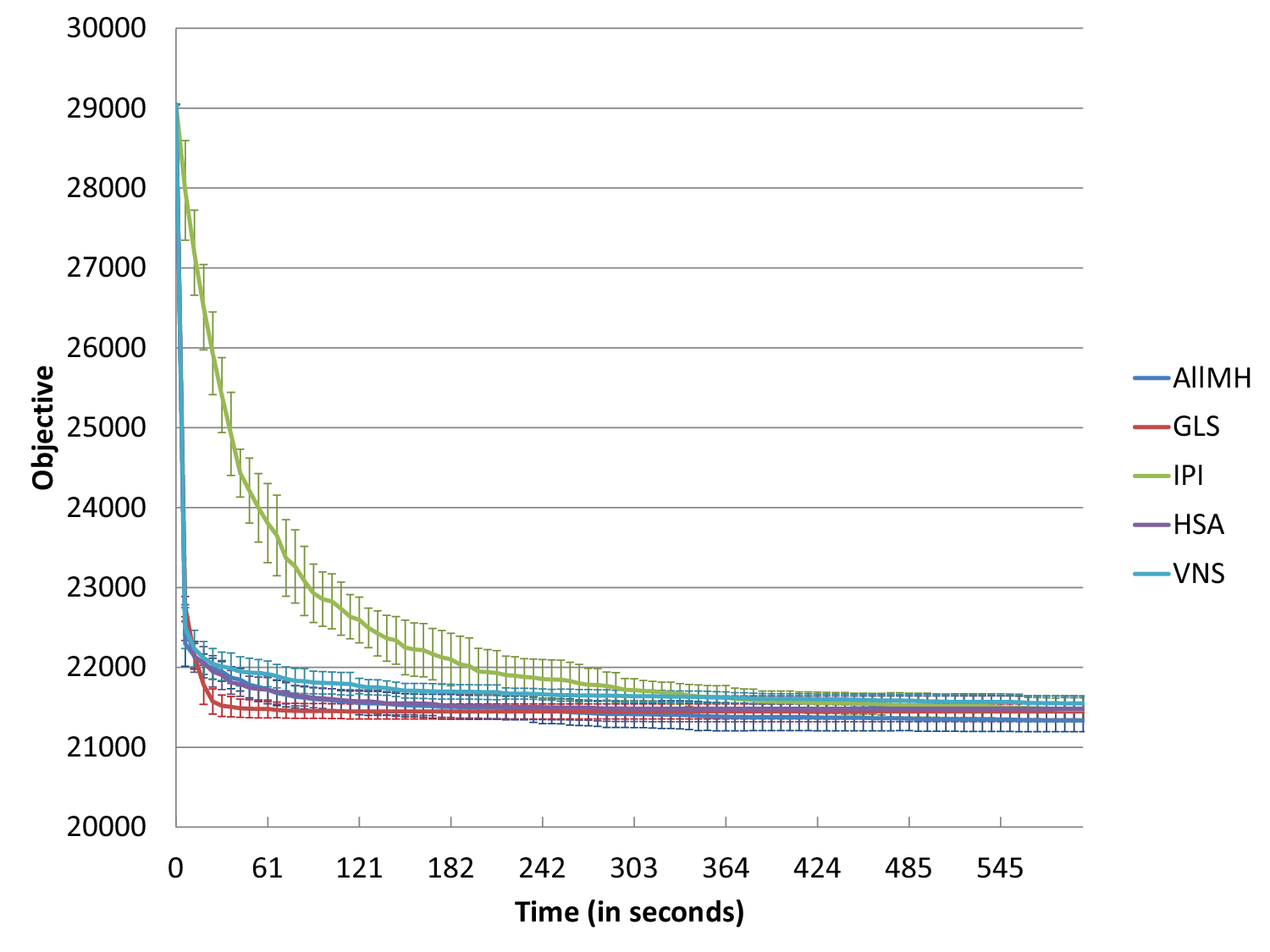}} \hfill
\subfloat[\label{graphicInd2}$n = 953,\:m=9350$]{\includegraphics[width=0.47\textwidth]{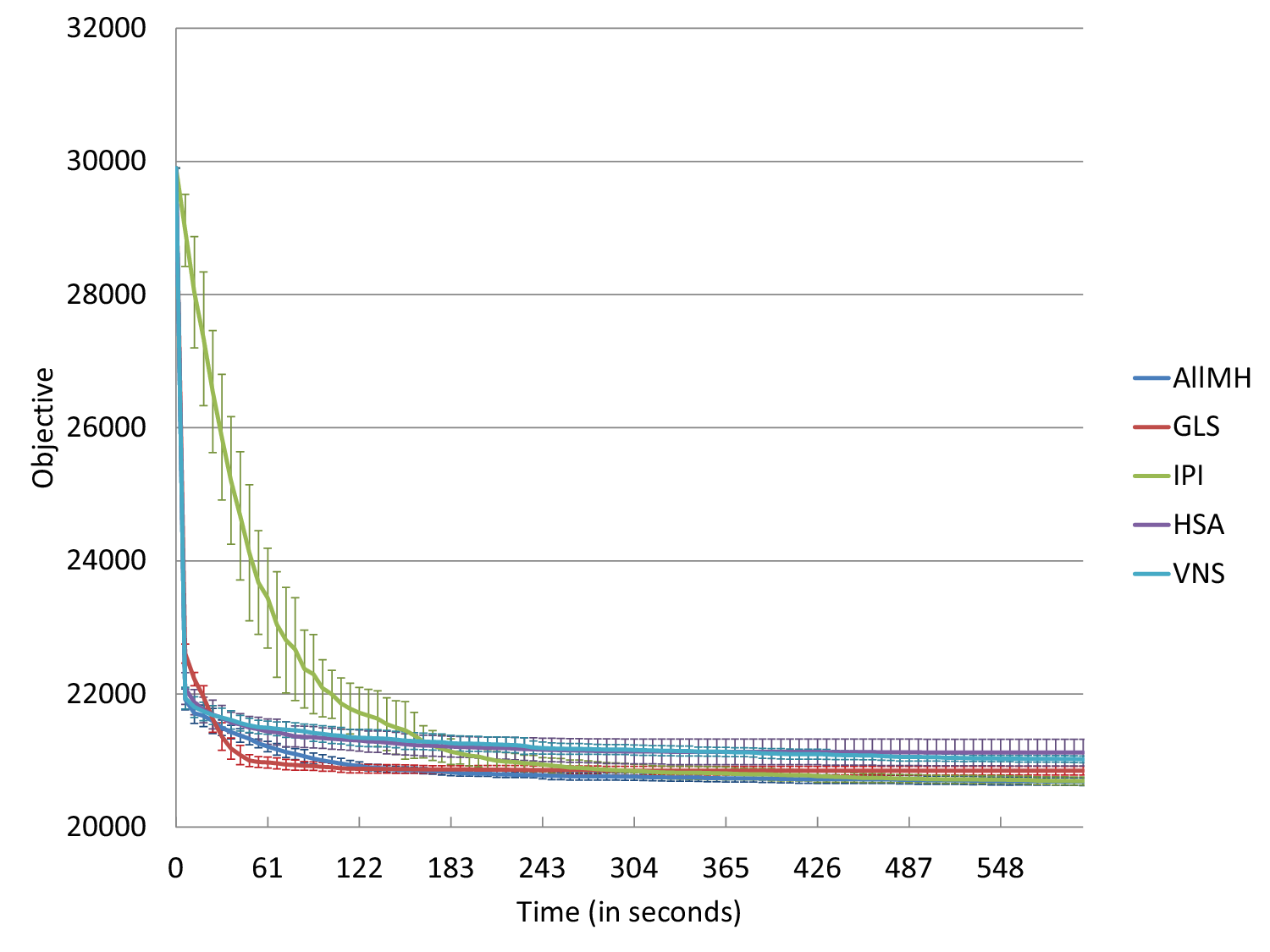}} \hfill
\caption{Industrial cases timeline of average objective values and standard deviations.} \label{graphicInd}
\end{figure}

For the meta-heuristics, it is useful to see the dynamic of improvement with time growth. We present the timeline for two industrial cases in Fig. \ref{graphicInd}. The first one with 688 vertices is given in Fig. \ref{graphicInd1}, and the second one with 953 vertices is given in Fig. \ref{graphicInd2}. In these figures, we show the average objective values by solid curves and the standard deviation values by vertical segments. It is seen that IPI improves solution slower than other algorithms. This is predictable behavior because IPI uses CPLEX, which is time consuming. Therefore, using of CPLEX is not reasonable for short-time runnings in such middle-sized cases. In both cases, GLS outperforms HSA and VNS within almost any time limit for the both cases. Combining all heuristics in the AllMH algorithm allows to find the best solution after 3-5 minutes, and during the first few minutes only GLS outperforms AllMH. In both cases, stabilization of all algorithms comes much earlier than 10 minutes pass.

\begin{figure}
\centering
\includegraphics[width=1\textwidth]{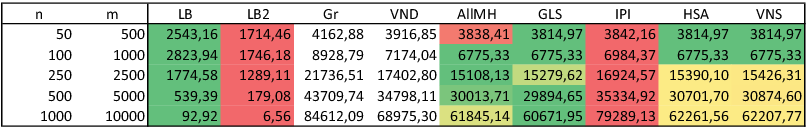}
\caption{Random graph. Average objective and lower bound values.} \label{tableRand}
\end{figure}

\begin{figure}
\centering
\includegraphics[width=0.8\textwidth]{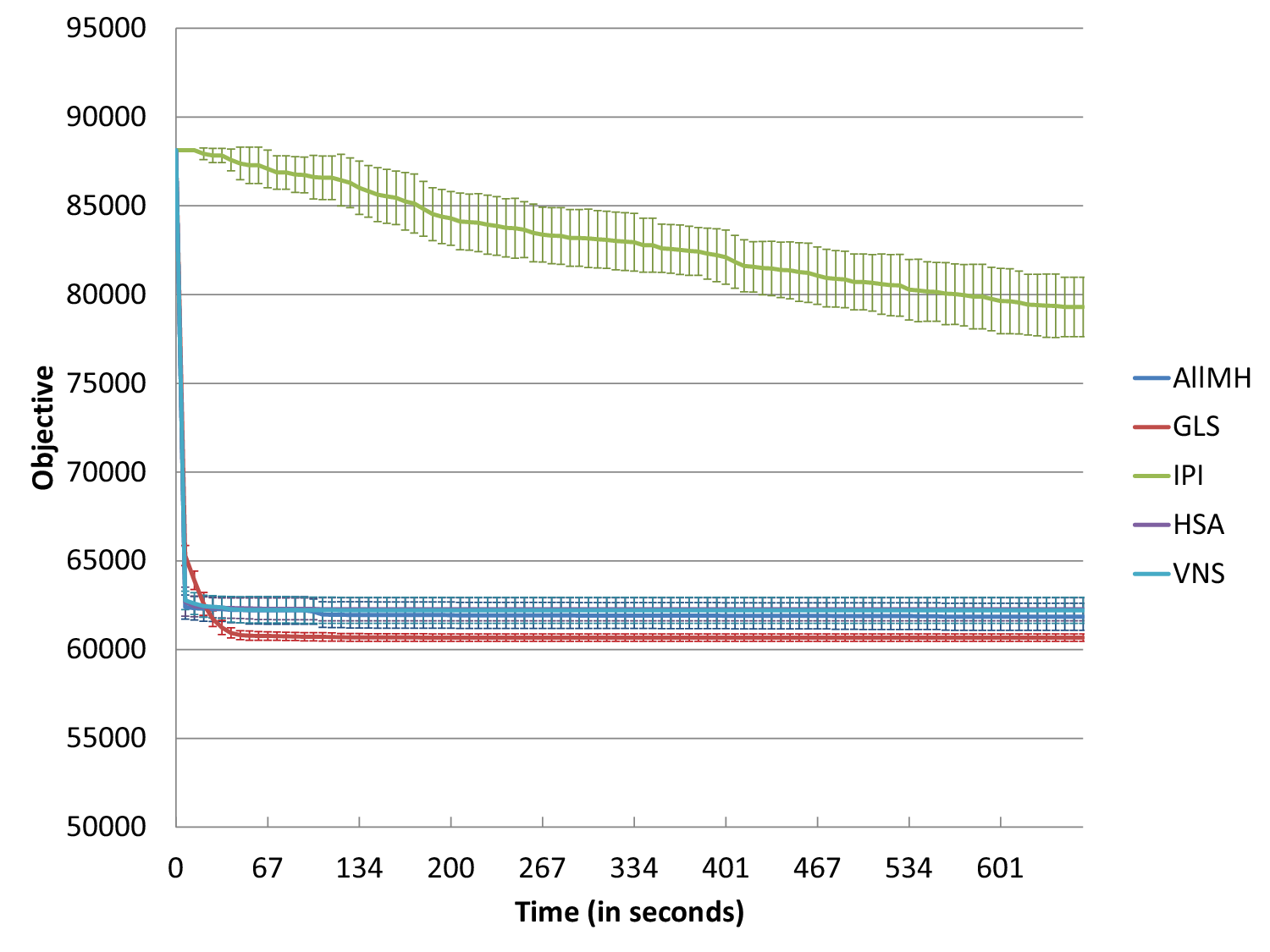}
\caption{Random graph. $n = 1000,\:m=10000$. Timeline of average objective values and standard deviations.} \label{graphicRand}
\end{figure}

The results for the random graph are presented in Fig. \ref{tableRand}. It is worth mentioning that IPI behaves worse than other algorithms this time. We believe that this happens because the components selected by IPI are not dense enough, and there remain more useful links between vertices of different components than for the graphs of the same size but less balanced vertices degrees values. This affects the calculation of the lower bounds, which appear to be too small for such graphs. The timeline for the case with 1000 vertices is presented in Fig. \ref{graphicRand}. It is seen that performance of IPI changes almost linearly regarding running time, while for all other methods a huge improvement comes by the end of the first minute, and then they stabilize.

\begin{figure}[!hbtp]
\centering
\subfloat[\label{tableUDG1}Average objective and lower bound values]{\includegraphics[width=1\textwidth]{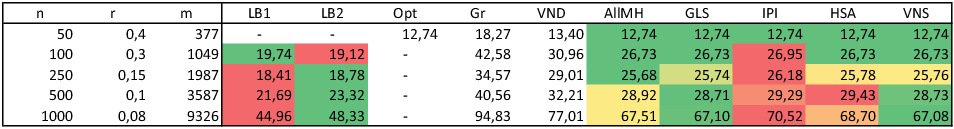}} \hfill
\subfloat[\label{tableUDG2}Ratio estimations]{\includegraphics[width=1\textwidth]{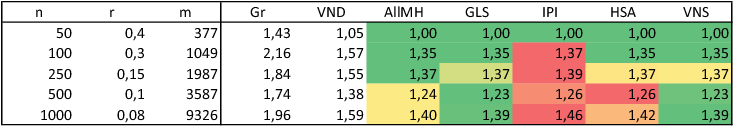}} \hfill
\caption{Unit disk graph results} \label{tableUDG}
\end{figure}

\begin{figure}
\centering
\includegraphics[width=0.8\textwidth]{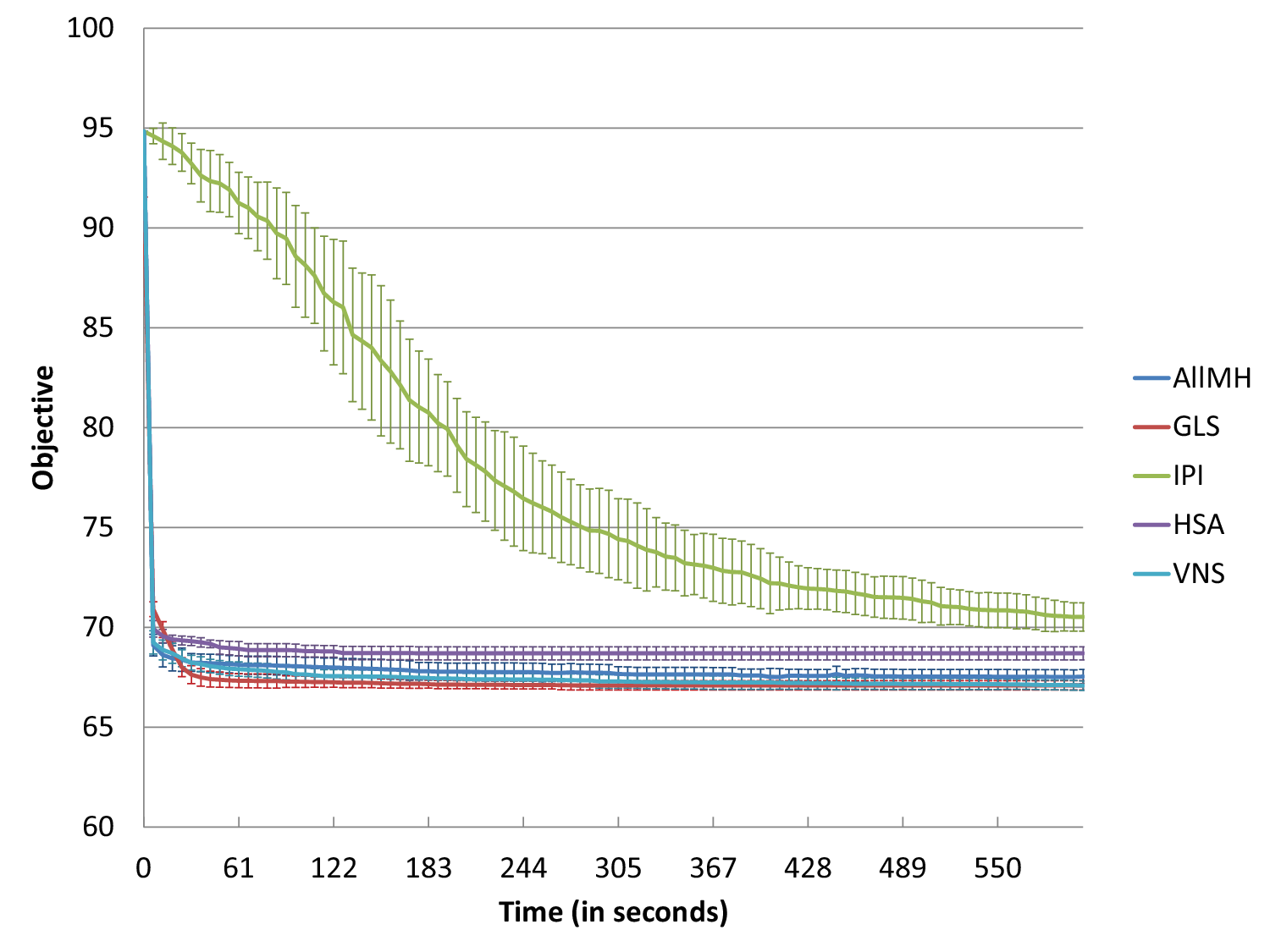}
\caption{Unit disk graph. $n = 1000,\:r=0.08,\:m=9326$. Timeline of average objective values and standard deviations.} \label{graphicUDG}
\end{figure}

The results for the unit disk graph are presented in Fig. \ref{tableUDG}. IPI yields the worst solution in almost all cases, while the best one is found by GLS with slight difference with other methods. The lower bounds and objective values of the best solutions are near optimal, and therefore the ratio is rather low (less than 1.5). The timeline for the case with 1000 vertices is given in Fig. \ref{graphicUDG}. In this figure, one may notice fast stabilization of all methods except IPI and almost parallel locations of their graphics: AllMH, VNS, and GLS give very close solutions during whole time period, HSA always goes with a significant gap between them, and IPI’s graphic lies much higher than HSA’s graphic so it gives much worse solution during whole time period, although the difference becomes less close to the end of this period.

We may conclude that GLS is more preferable than other meta-heuristics, but VNS and HSA build solutions with similar performance ratio and have easier implementations. IPI improves solution slower than other methods and works rather badly on randomly generated graphs, but for some particular industrial cases it outperforms other algorithms after a rather long period of running time. Combining different meta-heuristics in one algorithm AllMH appeared to be a powerful method that works well on all tested cases both in short and long running time periods.

\section{Conclusion}

This study addresses an NP-hard weighted graph 3-coloring problem. Instead of minimizing the number of colors while ensuring adjacent vertices have different colors—the traditional approach—our formulation uses edge-weighted graphs and assigns three colors to vertices to minimize the total weight of monochromatic edges. This optimization problem remains poorly explored in the literature.

Building upon prior work that introduced graph decomposition techniques and basic heuristic methods, we present novel approximation algorithms in this research. The proposed methods encompass spectral clustering based graph decomposition, meta-heuristic approaches (namely, variable neighborhood search, simulated annealing, and genetic algorithm), and utilization of CPLEX packet for solving the ILP and IQP formulations on the small-sized instances.

The experimental analysis reveals distinct advantages of the proposed optimization approaches. Genetic local search (GLS) demonstrates superior performance compared to alternative meta-heuristics. Variable neighborhood search (VNS) and hybrid simulated annealing (HSA) achieve comparable solution quality with notably simpler implementations. Iterative partial improvement (IPI) method exhibits slower convergence rates, showing significantly worse results on synthetic random graphs but occasionally delivering superior outcomes in specific industrial scenarios following extended computation periods. An effective technique emerged from the hybrid approach, integrating multiple meta-heuristics (AllMH). The AllMH demonstrates high efficiency, maintaining powerful performance across diverse test scenarios regardless of computation time.

\section*{Declarations}
\begin{itemize}
\item The manuscript was written on the initiative of the authors without financial support.
\item There are no conflicts of interest.
\item Ethical approval and consent for participation of the authors is available.
\item The authors agree to the publication of the manuscript.
\item All necessary data are in the text of the manuscript.
\item No additional materials are required.
\item The manuscript does not contain codes.
\item The authors' contribution is equal.
\end{itemize}

\end{document}